\documentclass[prl,twocolumn,showpacs,showkeys,preprintnumbers,amsmath]{revtex4}

\begin{document}
\preprint{IPM/P-2007/028 \cr
           \eprint{arXiv:0705.1879} }

\title{Lovelock Gravity at the Crossroads of Palatini and Metric
Formulations}

\author{Q. \surname{Exirifard}}
\affiliation{Institute for Studies in Theoretical Physics and Mathematics (IPM), P.O.Box 19395-5531, Tehran, Iran}

\email{exir, jabbari@theory.ipm.ac.ir}

\author{M. M. Sheikh-Jabbari}
\affiliation{Institute for Studies in Theoretical Physics and Mathematics (IPM), P.O.Box 19395-5531, Tehran, Iran}

\begin{abstract}
We consider  extensions of the Einstein-Hilbert Lagrangian to a
general functional of metric and  Riemann curvature tensor, ${\cal
L}(g_{\mu\nu}, R_{\mu\alpha\beta\nu})$. A given such Lagrangian
describes two different theories depending on considering connection
and metric (Palatini formulation), or only the metric (metric
formulation) as independent dynamical degrees of freedom.
Equivalence of the Palatini and metric formulations at the level of
equations of motion, which as we will argue is a manifestation of
the Equivalence Principle, is the physical criterion that restricts
 form of the Lagrangians of modified gravity theories. We prove that
within the class of modified gravity theories we consider, only the
Lovelock gravity satisfies this requirement.

\end{abstract}
\pacs{04.20.Fy, 04.50.+h, 04.60.-m}
\keywords{Lovelock gravity, Palatini formulation, Modified theories of gravity}
\maketitle

\section{Introduction}
General Relativity (GR) associates gravity to the (geometric)
properties of space-time, metric and  connection. These represent
two essentially different properties of space-time, metric is the
measure of  length  while  connection  defines the covariant
derivative and   parallel transportation. Geodesics, the curves
which extremize the distance between two points, are thus specified
with the metric. The worldline of a free particle, a curve along
which the velocity vector is covariantly constant,  is determined by
the connection. For a general connection the worldline of a free
particle need not be a geodesic.

In the ordinary \textit{metric formulation} of GR, we require that
the worldline of a free particle is a geodesic.  This requirement
fixes the connection to the Levi-Civita connection, the components
of which are  the  Christoffel symbols. Extremizing the action with
respect to the metric gives the equations of motion for the metric.
In principle one can relax this requirement, treat the connection
and  metric as two independent fields and  extremize the action with
respect to both to obtain respective equations of motion. We refer
to this latter treatment as the \textit{Palatini formulation}
\endnote[1]{There exists a debate in the literature on calling this
after Palatini \cite{Palatini-Form} or Einstein \cite{Einstein}.
Ref. \cite{Error} argues that it should be called after Einstein.}.
The connection solving the equation of motion of the connection in
the Palatini formulation does not necessarily coincide with the
Levi-Civita connection \footnote[2]{The inequivalence of the metric
and Palatini  (also known as metric-affine, first order or mixed)
formulations for a general Lagrangian is a well studied issue in the
literature; \textit{e.g.} \cite{non-Identical}.}. For the
Einstein-Hilbert action, however, only the Levi-Civita connection
solves the corresponding equation in the Palatini formulation and
the two formulations become identical \cite{{Palatini-Form},
{Einstein},{Error}}.

Although successful in describing the observational and the
experimental data, there exist theoretical and phenomenological
motivations to study modifications or corrections to  the
Einstein-Hilbert action. In the theoretical side, we know that the
Einstein-Hilbert action is a classical self-interacting theory. In
the semi-classical regime, in principle, this action receives
quantum corrections. For example string theory, as a candidate for
quantum gravity, provides  a framework for computing the higher
order corrections to the Einstein-Hilbert action up to the field
redefinition ambiguities \cite{ambiguities}. In a phenomenological
approach to cosmology and astrophysics, it has been argued that an
appropriate modification of the Einstein-Hilbert action may provide
an alternative resolution to dark matter and dark energy problems,
and  a natural framework to address the inflationary
paradigm~\footnote{Implications of $f(R)$ modified gravity theories
on cosmology, and astro and solar system physics bas been studied
extensively in the both Palatini and metric formulations;
\textit{e.g.} \cite{f(R)}.}.

In a bottom-up approach, the general covariance imposes a weak
restriction on the Lagrangian of  modified gravities. Therefore, it
is desirable to find additional theoretical criteria or requirements
to restrict further the form of the Lagrangian. We argue that the
equivalence of the Palatini and metric formulations, which is a
property of the Einstein gravity theory, naturally provides such
theoretical criterion and strongly restricts form of the corrections
or modifications to the Einstein-Hilbert action.

Let us  elaborate on the physical meaning of the equivalence of the
Palatini and metric formulations. As mentioned, in the Palatini
formulation a free particle does not necessarily follow a geodesic,
the path which minimizes the distance.  Consider a massless particle
(a light ray) which should follow a path of a free particle in a
given background geometry. If this path is not a geodesic, then
there should exist another path, a geodesic, along which an
(accelerated) object can travel  faster than light. This is in
contradiction with the basics of the Einstein general relativity. In
another point of view, along a geodesic the particle will feel a
force and hence gravity cannot be locally turned off. This is
against the usual interpretation of the Equivalence Principle. In
the metric formulation we do not face these contradictions.
Nonetheless, in a theory of modified gravity, there is always the
theoretical possibility of choosing the Palatini or metric
formulations and  there is no reason which one should be taken from
the outset.

We take the standpoint that the ``physically allowed'' theories of
the modified gravity are those for which the Palatini and metric
formulations are (classically) equivalent. Here we consider a class
of torsion-free modified gravity theories in which the gravity part
of the Lagrangian is only a functional of the Riemann tensor and
metric, and not of their covariant derivatives, and the matter part
of the Lagrangian does not involve the connection. We prove that
within this class only the Lovelock gravity theories fulfill the
requirement of equivalence of the Palatini and metric formulations.

To this end we take the following route. Deriving the equations of
motion for a general Lagrangian of interest in both the metric and
Palatini formulations, we first require the \textit{consistency} of
the two formulations. We implement the consistency by demanding that
the Levi-Civita connection solves the equation of motion of the
connection in the Palatini formulation. This makes the equation of
motion for the metric identical in the both formulations. We show
that only the Lovelock gravity meets the consistency requirement.
 We then prove the \textit{equivalence} of the two
formulations for the Lovelock gravity by considering the Lovelock
theory in the Palatini formulation. We show that in the
asymptotically flat space-times, only the Levi-Civita connection
solves the equations of motion for the connection.

\section{Palatini Vs. metric formulations}
We consider the modified  gravity Lagrangians of the following form
(where no explicit covariant derivative is involved): \small
\begin{equation}\label{modified-action}
    S_{mod. GR}=\frac{1}{4\pi G_N}\int d^Dx\ \sqrt{-g}\
    {\cal L}(g_{\mu\nu}, R_{\mu \nu\alpha\beta})\,,
\end{equation}%
\normalsize%
where the Riemann curvature tensor is defined by :%
\small
\begin{equation}\label{Riemann-def}\begin{split}%
R_{\ \beta\mu\nu}^{\alpha} &\equiv 2
\partial_{[\mu}\Gamma^\alpha_{\nu]\beta}+2
\Gamma^\alpha_{[\mu|\rho}\Gamma^\rho_{\nu]\beta},\\
R_{\alpha \mu\beta\nu} &\equiv g_{\alpha\eta} R_{\ \mu\beta\nu}^{\eta}\,,%
\end{split}\end{equation}%
\normalsize%
and $\Gamma^\alpha_{\beta\nu}$ is the connection. In this work we
consider the \textit{torsion-free} theories,
$\Gamma^\alpha_{\beta\nu} = \Gamma^\alpha_{\nu\beta}$. We assume
that the matter part of the Lagrangian does not contain the
connection. The whole action (the matter plus the gravity parts) can
be understood either in the \textit{metric formulation} where the
connection is the Levi-Civita connection, or in the \textit{Palatini
formulation} where the connection is an independent variable and
does not necessarily coincide with the Levi-Civita.

Let us derive the equations of motion of this action in both the
metric and Palatini formulations. The equations of motion in the
metric formulation are \cite{Wald}%
 \small
\begin{subequations}\label{METRIC-EQUATIONS}
\begin{align}\label{Levi-Civita}%
&\Gamma^{\mu}_{\alpha\beta}=\{^\mu_{\alpha\beta}\}=\frac{1}{2}g^{\mu\nu}(\partial_\alpha
g_{\beta\nu}+\partial_\beta g_{\nu\alpha}-\partial_\nu
g_{\alpha\beta})\,,\\
&\label{metric-eom}
\frac{\partial {\cal L}}{\partial g_{\mu\nu}}
+\frac{1}{2} {\cal L} g^{\mu\nu}
+\frac{\partial {\cal L}}
    {\partial R_{\mu\rho\alpha\beta}}R_{\ \rho\alpha\beta}^{\nu}%
+2\nabla_{\{\alpha}\nabla_{\beta\}}
    \frac{\partial{\cal L}}{\partial R_{\mu\alpha\beta\nu}}
= -T^{\mu\nu},
\end{align}
\end{subequations}
\normalsize %
where $\{^{\mu}_{\alpha\beta}\}$ is the Christoffel
symbol, while in the Palatini formulation the equations of motion
for the connection and the metric read \small
\begin{subequations}\label{Palatini-eom}
\begin{align}\label{Palatini-eom-c}
\nabla_\nu
    (
    \sqrt{-g}
    \frac{\partial{\cal L}}{\partial R_{\mu\{\alpha\beta\}\nu}}
    g_{\mu\rho}
    )&=0, \\ \label{Palatini-eom-m}
\frac{\partial {\cal L}}{\partial g_{\mu\nu}}
+\frac{1}{2} {\cal L} g^{\mu\nu}
+ \frac{\partial {\cal L}}
    {\partial R_{\mu\rho\alpha\beta}}R_{\ \rho\alpha\beta}^{\nu}
&=-T^{\mu\nu},
\end{align}
\end{subequations}
\normalsize %
where $T^{\mu\nu}$ in the r.h.s of \eqref{metric-eom} and
(\ref{Palatini-eom}b) stands for the energy-momentum tensor of the
matter field. Note that the partial derivatives of ${\cal L}$ are
taken assuming that $g_{\mu\nu}$ and $R_{\mu\nu\alpha\beta}$ are
independent variables, and the partial derivative coefficients are
uniquely fixed to have precisely the same tensor symmetries as the
varied quantities. For a general Lagrangian,
\eqref{METRIC-EQUATIONS} and \eqref{Palatini-eom} are  not
equivalent \cite{endnote2} or even consistent.

\section{Requiring the consistency}

Requiring the consistency of the two formulations amounts to
demanding  (\ref{METRIC-EQUATIONS}a) or equivalently,
\begin{equation}\label{metric-compatible}%
    \nabla_\alpha g_{\mu\nu}=0\,,
\end{equation}
to be a solution of (\ref{Palatini-eom}a). Expand
(\ref{Palatini-eom}a) as follows%
 \small
\begin{eqnarray}\label{expand-Eq1}
\frac{\partial{\cal L}}{\partial R_{\mu\{\alpha\beta\}\nu}}
\nabla_\nu
    (
    \sqrt{-g} g_{\mu\rho})
+\sqrt{-g} g_{\mu\rho}\nabla_\nu
(
    \frac{\partial{\cal L}}{\partial R_{\mu\{\alpha\beta\}\nu}}
)&=0\,.
\end{eqnarray}
\normalsize%
When \eqref{metric-compatible} holds the first term of
\eqref{expand-Eq1} vanishes yielding %
 \small
\begin{equation}\label{expand-Eq2}
 \nabla_\nu
(
    \frac{\partial{\cal L}}{\partial R_{\mu\{\alpha\beta\}\nu}}
)=0\,.
\end{equation}
\normalsize%
Now let ${\cal L}$ have a Taylor expansion in terms of the Riemann
tensor. For Lagrangians of our interest the expansion coefficients
are functionals of metric only and when \eqref{metric-compatible}
holds these coefficients are covariantly constant. Therefore, in our
case we can use the ``chain rule with the covariant derivative'' on \eqref{expand-Eq2} to obtain%
\begin{equation}\label{Gamma-compatible}%
\frac{\partial^2{\cal L}}{\partial R_{\mu\{\nu\beta\}\alpha}\partial
R_{\rho\sigma\lambda\gamma}}\nabla_\alpha
R_{\rho\sigma\lambda\gamma}=0.%
\end{equation}
Recalling the Bianchi identity $\nabla_{[\alpha}
R_{\rho\sigma]\lambda\gamma}=0$, \eqref{Gamma-compatible} is
guaranteed to be satisfied if~\footnote{Note that with the
Levi-Civita connection, \emph{i.e.} when \eqref{metric-compatible}
holds, the Riemann tensor has its usual
 symmetries on its indices, \textit{e.g.}
$R_{\lambda\gamma\rho\sigma} = R_{\rho\sigma\lambda\gamma}$.}
\small
\begin{equation}\label{Cyclic-derivative}%
\frac{\partial^2{\cal L}}
{\partial R_{\mu\{\nu\beta\}\alpha}\partial R_{\lambda\gamma\rho\sigma}}
= \frac{\partial^2{\cal L}}
{\partial R_{\mu\{\nu\beta\}\sigma}\partial R_{\lambda\gamma\alpha\rho}}
= \frac{\partial^2{\cal L}}
{\partial R_{\mu\{\nu\beta\}\rho}\partial R_{\lambda\gamma\sigma\alpha}},
\end{equation}%
\normalsize%

In what follows we  show that only the Lovelock gravity Lagrangian
satisfy the Palatini-metric consistency requirement, summarized in
\eqref{Cyclic-derivative}. Let us, however, first briefly review the
Lovelock theory. David Lovelock used the following assumptions to
restrict form of the action for pure gravity including the higher
derivative terms \cite{Lovelock}:
\begin{enumerate}%
\item The generalization of the Einstein tensor, hereafter denoted by $A_{\mu\nu}$,
should be a symmetric tensor of rank two; $A_{\mu\nu}=A_{\nu\mu}$,

\item $A_{\mu\nu}$ is concomitant of the metric and its first two derivatives,
$A_{\mu\nu}\,=\,A_{\mu\nu}(g,\partial g,\partial^2g)$,

\item $A_{\mu\nu}$ is divergence free, $\nabla^\mu A_{\mu\nu}\,=\,0$.
\end{enumerate}
(We should stress that Lovelock was working in the ``metric
formulation'' assuming $\nabla_\alpha g_{\mu\nu}=0$.) In a series of
theorems \cite{Lovelock2}, Lovelock proved that the above three
assumptions are fulfilled for the generalized Einstein tensor
derived only from the following Lagrangian density
\cite{Lovelock-Actions}
\small\begin{equation}\label{LLL}%
 {\cal
L}_{Lovelock}=\sum_{n=0}^{[\frac{D}{2}]} a_{n}\,
\theta^{\mu_1\cdots \mu_{2n} \nu_1 \cdots \nu_{2n}}(g)
\prod_{p=1}^{n}\, R_{\mu_{2p-1} \mu_{2p} \nu_{2p-1} \nu_{2p}} \,,
\end{equation}
\normalsize where $D$ is the number of the dimensions of
space-time and $[\frac{D}{2}]$ represents the integer part of
$\frac{D}{2}$, and
\begin{eqnarray}\label{delta}
\theta^{ \mu_1\cdots \mu_{2n} \nu_1 \cdots \nu_{2n}}(g) &=&\det
\left|
\begin{array}{lll}
g^{\mu_1 \nu_1} & \cdots & g^{\mu_{2n} \nu_1}\\
\vdots&&\vdots\\
g^{\mu_1 \nu_{2n}}&\cdots& g^{\mu_{2n} \nu_{2n}}
\end{array}\right|\,
\end{eqnarray}
$a_n$'s are some constant values of proper dimensionality. We refer
to the $n^{th}$ term in \eqref{LLL} as the $n^{th}$ order Lovelock
gravity. At its zeroth and first order, Lovelock gravity coincides
respectively with the cosmological constant and the Einstein-Hilbert
action. Its second order coincides with the Gauss-Bonnet Lagrangian
density. The compact form of its higher orders becomes more
involved, e.g. see~\cite{muller,wheeler} for the explicit form of
the third and fourth orders~\footnote{Note that except for $n=1$,
none of the Lovelock gravities are of the form of ${\cal L}={\cal L}
(R, R_{\mu\nu})$.}.

In order to prove that the consistency of  the Palatini and metric
formulations happens only for the Lovelock gravities,  we first show
that when the consistency holds  the generalized Einstein tensor
satisfies the Lovelock's assumptions.  The generalized Einstein
tensor is defined by the variation of the action with respect to
metric, that is
\begin{subequations}\label{Ein-tensor-action}
\begin{align} A_{\text{metric}}^{\mu\nu}&\equiv
 \text{The l.h.s of the Eq. \eqref{metric-eom}}\,, \\
A_{\text{Palatini}}^{\mu\nu} &\equiv \text{The l.h.s of the Eq.
(\ref{Palatini-eom}b)}\,,
\end{align}\end{subequations}
where the subscripts indicates the formulation. The Palatini-metric
consistency requirement  implies that
$A_{\text{metric}}^{\mu\nu}=A^{\mu\nu}_{\text{Palatini}}$, which
using their explicit forms  presented in \eqref{metric-eom} and
(\ref{Palatini-eom}b), can be expressed as%
\small
\begin{equation}\label{Palat-Ein-tensor1}
A_{\text{metric}}^{\mu\nu} - A^{\mu\nu}_{\text{Palatini}}= 2
\nabla_{\{\alpha}\nabla_{\beta\}} \frac{\partial{\cal L}}{\partial
R_{\mu\alpha\beta\nu}}=0\,.
\end{equation}
\normalsize %
Assuming \eqref{metric-compatible} and \eqref{expand-Eq2}, and
recalling the symmetries of the Riemann tensor, one can immediately
verify that the above equation holds. The remaining step is to
recall that the Riemann tensor --due to \eqref{metric-compatible}--
contains at most the second  derivative of the metric and
$A_{Palatini}^{\mu\nu}$, by definition, involves the powers of the
Riemann tensor and not its derivatives. Therefore, the generalized
Einstein tensor
 being concomitant of the metric and its first two
 derivatives fulfills the second Lovelock assumption. Since
we have derived the Einstein tensor from a generally covariant
Lagrangian,  the first and third Lovelock assumptions hold too. The
uniqueness theorems of Lovelock \cite{Lovelock2} then proves that
the consistency of the metric and Palatini formulation can hold only
for the Lovelock gravities.

In order to complete the proof of the consistency, we must verify
that  for Lovelock Lagrangians, \eqref{metric-compatible} solves
either the equation of motion for the connection
(\ref{Palatini-eom}a), or  equivalently \eqref{Cyclic-derivative}.
This verification is obvious if we insert \eqref{LLL} into
\eqref{Cyclic-derivative} and recall the invariance of the
determinant \eqref{delta} under the cyclic permutation of its three
rows. (Lovelock Lagrangians satisfy \eqref{Cyclic-derivative}
without symmetrization  over $\beta$ and $\nu$ indices.)

\section{The equivalence of the formulations}

To argue for the equivalence, we should show that the only solution
to (\ref{Palatini-eom}a) for ${\cal L}={\cal L}_{Lovelock}$ is the
Levi-Civita connection. To this end,  we notice that when
$\nabla_\alpha g_{\mu\nu}\neq 0$, generically $R_{\mu\nu\eta\gamma}
\neq R_{\eta\gamma\mu\nu}$. However,  we note that
\begin{center}
%\begin{lemma} \label{LLL-property1}
\textit{for a general connection, the Lovelock Lagrangians are only
functional of the part of the Riemann tensor which  satisfies
$R_{\nu\mu\eta\gamma} = - R_{\mu\nu\eta\gamma}$ and
$R_{\mu\nu\eta\gamma} = R_{\eta\gamma\mu\nu}$.}
\end{center}
%\end{lemma}
This follows from the definition of the Lovelock lagrangian
densities in  \eqref{LLL}, and the antisymmetric property of the
determinant \eqref{delta} under permutation of its two rows or
columns, and the Bianchi identity of  $R_{\mu[\nu\eta\gamma]} = 0$.

The above lemma then implies that the Lovelock Lagrangians satisfy
\eqref{Gamma-compatible} and \eqref{Cyclic-derivative}, and
subsequently  \eqref{expand-Eq2}, for a general connection.
Therefore, the equation of motion of the connection
in the Palatini
formulation \eqref{expand-Eq1} takes the form%
 \small
\begin{equation}\label{Wha}
\frac{\partial{\cal L}_{\text{Lovelock}}}{\partial R_{\mu\{\alpha\beta\}\nu}}
\frac{\nabla_\nu(\sqrt{-g} g_{\mu\rho})} {\sqrt{-g}}
%&+&\\\nonumber
+ g_{\mu\rho}
    \frac{\partial^2{\cal L}_{\text{Lovelock}}}
    {\partial R_{\mu\{\alpha\beta\}\nu}\partial g_{ab}}
    \nabla_\nu g_{ab}=0\ .
\end{equation}
\normalsize
%where the successive use of the Leibniz rule is understood on \eqref{LLL}.
For the first order Lovelock Lagrangian density, \eqref{Wha} is an
algebraic equation for the connection whose unique solution is the
Levi-Cevita connection (see \textit{e.g.} section 3.4 of
\cite{Ortin}). For a general Lovelock Lagrangian density in the
Palatini formulation, however, \eqref{Wha} is a first order
differential equation for the connection.  If the connection
coincides with the Levi-Cevita connection in a single point on a
regular and connected space-time manifold then the uniqueness
theorems of the solutions to the differential equations guarantee
that the connection is the Levi-Cevita connection globally.

In an asymptotically flat space-time, the Riemann tensor vanishes in
the asymptotic infinity. Since both of the Riemann tensor and the
torsion vanish in the asymptotic infinity then we can choose
coordinates in such a way that the connection coincides with the
Levi-Cevita connection in the point of the asymptotic infinity.
Therefore, in the asymptotically flat space-times, the equivalence
of the Palatini and metric formulation for the Lovelock gravity is
guaranteed.

\section{Summary and Outlook}
%We have discussed that the equivalence of the Palatini and metric
%formulations is a criterion restricting the possible forms of the
%higher derivative corrections to gravity. We have inspired physical
%intuitions supporting this criterion. We have shown, provided the
%torsion vanishes and the Lagrangian is not a functional of the
%covariant derivatives of the curvature or metric, that in an
%asymptotically flat space-time, only the Lovelock gravity holds the
%equivalence

We have discussed that a strict interpretation of the Equivalence
Principle, in the absence of torsion, requires the connection to be
the Levi-Civita connection.  When the Lagrangian is not a functional
of covariant derivatives of the curvature or metric, this
requirement implies \textit{the consistency of Palatini and metric
formulations} and restricts the Lagrangians only to the Lovelock
gravity~\footnote{Similarities between the Einstein-Hilbert action
and the Lovelock gravity  have also been noted in \cite{Padi}.}.

The requirement of equivalence or consistency  of the Palatini and
metric formulations can be imposed on more general theories than
those we considered here. For example one may use this requirement
to restrict the form of action for gravity when torsion does not
vanish, or when the Lagrangian involves the covariant derivatives of
Riemann or metric. It can also be used to restrict form of the
non-minimal coupling between matters and gravity \cite{progress}.

The equivalence of the Palatini and metric formulations can also
serve as a criterion for fixing the field redefinition ambiguities
arising in the string loop or worldsheet corrections to
supergravities \cite{ambiguities}.  The proposals for fixing the
field redefinition ambiguities include the MM-criterion
\cite{MMcriterion} and the ghost-free condition \cite{ghost-free}.
Noting that Lovelock Lagrangians are ghost-free \cite{Zumino},  the
Palatini-metric equivalence criterion is in agreement with the
ghost-free criterion. Investigating the  Palatini-metric equivalence
criterion for matter fields non-minimally coupled to gravity,
however, precedes its comparison to the MM-criterion.


\begin{thebibliography}{9}
\bibitem{Palatini-Form}
A.~Palatini,
%   \textit{Deduzione invariantiva delle equazioni gravitazionali dal. principio di Hamilton},
    Rend. Circ. Mat. Palermo \textbf{43} (1919) 203.
\bibitem{Einstein}
A. Einstein, {Sitzung-ber Preuss Akad. Wiss.}, (1925)414.

\bibitem{Error}
M. Ferraris \textit{et al.},
%, M. Francaviglia and C. Reina,
%   \textit{Error},
    Gen. Rel. Grav. \textbf{14} (1981) 243.

\bibitem{non-Identical}
M. Ferraris  \textit{et al.},
%, M.~Francaviglia and I.~Volovich,
    %\textit{ The universality of vacuum {E}instein equations with cosmological constant,}
    Class. Quant. Grav.  \textbf{11} (1994) 1505;
%%CITATION = CQGRD,11,1505;%%
%
F. W. Hehl  \textit{et al.},
%, J.~D.~McCrea, E.~W.~Mielke and Y.~Neeman,
    %\textit{ Metric affine gauge theory of gravity: Field equations, Noether identities, world spinors, and breaking of dilation invariance,}
    Phys. Rept.  \textbf{258} (1995) 1;
%,
%    [arXiv:gr-qc/9402012].
    %%CITATION = PRPLC,258,1%%
T. P. Sotiriou  \textit{et al.},
%, and S.~Liberati,
    %\textit{ Metric-affine $f(R)$ theories of gravity,}
    Annals Phys.  \textbf{322} (2007) 935;
%    [arXiv:gr-qc/0604006].
%%CITATION = APNYA,322,935;%%
H. Weyl, %\textit{ A remark on the coupling of gravitation and electron,}
  Phys. Rev.  \textbf{77} (1950) 699.
%%CITATION = PHRVA,77,699;%%

\bibitem{Ortin}
T. Ortin,
    \textit{Gravity and strings},
    %\href{http://www.slac.stanford.edu/spires/find/hep/www?irn=5854970}{SPIRESentry}
    {Cambridge Unversity, Cambridge University Press, 2004.}

\bibitem{ambiguities}
A. A. Tseytlin,
    %\textit{ Ambiguity in the effective action in string theories,}
    Phys. Lett.  \textbf{B176} (1986) 92.
    %%CITATION = PHLTA,B176,92;%%

\bibitem{f(R)}
M. Amarzguioui \textit{et al.},%, O.~Elgaroy, D.~F.~Mota and T.~Multamaki,
    %\textit{  Cosmological constraints on $f(R)$ gravity theories within the Palatini  approach,}
Astron. Astrophys.  \textbf{454}(2006) 707;
%    [arXiv:astro-ph/0510519].
    %%CITATION = AAEJA,454,707;%%
R. Bean  \textit{et al.},
%, D.~Bernat, L.~Pogosian, A.~Silvestri and M.~Trodden,
    %\textit{  Dynamics of linear perturbations in f(R) gravity,}
    Phys. Rev.  \textbf{D75}(2007) 064020;
%    [arXiv:astro-ph/0611321].
    %%CITATION = PHRVA,D75,064020;%%
S. M. Carroll  \textit{et al.},
% ,A.~De Felice, V.~Duvvuri, D.~A.~Easson, M.~Trodden and M.~S.~Turner,
    %\textit{ The cosmology of generalized modified gravity models,}
    Phys. Rev.  \textbf{D71} (2005) 063513;
%   [arXiv:astro-ph/0410031].
    %%CITATION = PHRVA,D71,063513;%%
T. Chiba  \textit{et al.},
%, T.~L.~Smith and A.~L.~Erickcek,
    % \textit{  Solar system constraints to general f(R) gravity,}
    arXiv:astro-ph/0611867;
    %%CITATION = ASTRO-PH/0611867;%%
T. Koivisto  \textit{et al.},
%, and H.~Kurki-Suonio,
    %\textit{  Cosmological perturbations in the Palatini formulation of modified gravity,}
    Class. Quant. Grav.  \textbf{23} (2006) 2355;
%    [arXiv:astro-ph/0509422].
    %%CITATION = CQGRD,23,2355;%%
S. Nojiri \textit{et al.},
%and S.~D.~Odintsov,
    %\textit{   Introduction to modified gravity and gravitational alternative for dark energy,}
    Int. J. Geom. Meth. Mod. Phys. \textbf{4} (2007) 115.
%    [arXiv:hep-th/0601213].
    %%CITATION = 00436,4,115;%%

\bibitem{Wald}
  V. Iyer  \textit{et al.},
%and R.~M.~Wald,
  %``Some properties of Noether charge and a proposal for dynamical black hole
  %entropy,''
  Phys. Rev.  {\bf D50}, 846 (1994)
%  [arXiv:gr-qc/9403028].
  %%CITATION = PHRVA,D50,846;%%


\bibitem{Lovelock}
D. Lovelock,
    %\textit{  The Einstein tensor and its generalization},
    J. Math. Phys. \textbf{12} (1971) 498.

\bibitem{Lovelock2}
D. Lovelock,
    %\textit{  Divergence-free tensorial concomitants},
    Aequationes Math. \textbf{4} (1970) 127.





\bibitem{Padi}
A. Mukhopadhyay  \textit{et al.},
%and T.~Padmanabhan,
    %\textit{  Holography of gravitational action functionals,}
    Phys. Rev.  \textbf{D74} (2006) 124023.
%    [arXiv:hep-th/0608120].
    %%CITATION = PHRVA,D74,124023;%%

\bibitem{Lovelock-Actions}
D. Lovelock,
    \textit{Tensors differential forms and variational principles},
    Wiley-Interscience, New York, 1975.


\bibitem{muller}
F. Mueller-Hoissen,
    %\textit{  Spontaneous compactification with quadratic and cubic curvature terms},
    Phys. Lett. \textbf{B163} (1985) 106.

\bibitem{wheeler}
J. T. Wheeler,
    %\textit{  Extended Einstein equations},
    Nucl. Phys. \textbf{B268} (1986) 737.

\bibitem{progress}
M.M. Sheikh-Jabbari and A. Vahedi,
    \textit{in progress}.

\bibitem{MMcriterion}
I. Jack  \textit{et al.},
%and D.~Jones,
    %\textit{  $\sigma$-model $\beta$-functions and
    %ghost free  string effective actions},
    Nucl. Phys. \textbf{B303} (1988) 260.

N. Mavromates  \textit{et al.},
%and J.~Miramontes,
    %\textit{  Effective actions from the conformal
    %  invariance conditions of bosonic $\sigma$ models with graviton and dilaton background},
    Phys. Lett. \textbf{B201} (1988) 473.

\bibitem{ghost-free}
D. Blas,
    %\textit{  Gauge symmetry and consistent spin-two theories,}
    arXiv:hep-th/0701049.
    %%CITATION = HEP-TH/0701049;%%

I. Jack  \textit{et al.},
%, D.~R.~T.~Jones and A.~M.~Lawrence,
    % \textit{  Ghost freedom and string actions,}
    Phys. Lett. \textbf{B203} (1988) 378.
    %%CITATION = PHLTA,B203,378;%%

B. Zwiebach,
    % \textit{  Curvature squared terms and string theories,}
    Phys. Lett. \textbf{B156} (1985) 315.
     %%CITATION = PHLTA,B156,315;%%
\bibitem{Zumino}
  B.~Zumino,
  %``Gravity Theories In More Than Four-Dimensions,''
  Phys.\ Rept.\  {\bf 137} (1986) 109.
  %%CITATION = PRPLC,137,109;%%



\end{thebibliography}
\end{document}